\documentstyle[twocolumn,aps]{revtex}

\begin{document}

\title{Radial fluctuations induced stabilization of the ordered state in two 
dimensional classical clusters.}

\author{I. V. Schweigert \cite {*:gnu},
V. A. Schweigert \cite {*:gnu} and   F. M.  Peeters \cite {f:gnu}}
\address{\it  Departement Natuurkunde, Universiteit 
Antwerpen (UIA),\\
Universiteitsplein 1, B-2610 Antwerpen, Belgium}
\date{\today}
\maketitle 

\begin{abstract}
Melting of two dimensional (2D) clusters of classical particles is studied using 
Brownian  dynamics and Langevin molecular dynamics simulations. The particles 
are confined by a circular 
hard wall or a parabolic external potential and interact through a dipole or a 
screened Coulomb potential. We found that with decreasing strength of the 
inter--particle interaction clusters with short-range inter-particle interaction 
which are confined by a hard wall exhibit a  
re-entrant behavior in its orientational order. 
\end{abstract}

The structural and dynamical properties of small classical two-dimensional (2D)
clusters have been the subject of recent 
experimental studies \cite {Lee,Mitchell,Bubeck} and Monte-Carlo and 
molecular dynamics 
simulations \cite 
{Lozovik90,Beda94,Schweigert95,Schweigert96,Schweigert98,Belousov98,candido}. It was 
found earlier that the particles are arranged in shells and that melting 
of finite clusters is a two step process. With increasing 
temperature inter-shell motion develops and the system losses  angular order. 
Consecutively radial 
diffusion switches on and destroys the shell structure of the cluster.  
The spectrum of such clusters was obtained in Refs.\cite 
{Schweigert95,Schweigert96,Belousov98} for different number of particles.
The derived minimal 
frequencies and the corresponding energy barriers showed that `non-close 
packed' clusters are unstable against inter-shell rotation. High symmetry 
clusters (i.e. the so-called {\it magic number} clusters) have energy barriers 
for intershell motion which are several orders of magnitude larger 
than those for `non-close packed' clusters.

Recently, Bubeck {\it et al}\cite {Bubeck} observed re-entrant melting in two 
dimensional (2D) colloidal clusters. The clusters consist of   
paramagnetic colloidal spheres which were confined in a circular hard wall
vessel. 
The external magnetic field induces a magnetic moment $\vec M$ in the particles 
and they interact through a dipole potential
$V ({\vec {r}}_i,{\vec {r}}_j)=\mu _0M^2/4\pi r_{ij}^3$, where 
$\mu_0$ is the magnetic permitivity, and $r_{ij}$ the interparticle distance.
The coupling parameter, which is the inter-particle interaction energy  
measured in units of 
the particle kinetic energy $\Gamma =V/k_BT$, characterizes the order of 
the system. It decreases by lowering the external magnetic field.
In Ref. \cite{Bubeck} it was found that with decreasing $\Gamma $, 
first inter-shell rotation appears which destroys the angular order of 
the cluster. Further decreasing the parameter $\Gamma $ the 
system unexpectantly regained 
angular order within a narrow range of $\Gamma $ and then melts when
$\Gamma$ is further decreased. It was suggested that the 
observed re-entrant melting behavior  
is due to the increasing role of the radial 
particle fluctuations which is very similar to an earlier investigation of 
laser induced melting of 2D colloidal crystals \cite {Chakrabarti,Wei}. 

Earlier theoretical work \cite{Beda94}
on parabolic confined clusters did not find such a 
re-entrant behavior which suggests that the shape of the confinement
potential may be very important.  
Another difference is that in the experimental system the particle motion
is strongly damped because the colloidal particles move in water.  
In the present paper we investigate the mechanism for the re-entrant behavior
and address the specific role played by the type of confinement (i.e. hard wall
versus parabolic)
and of the functional form of the inter-particle interaction (short range
versus long range) on the melting of the clusters. 

In our model the particles are confined by a circular hard wall potential ($V_p=0$ 
for $r\leq R$ and $V_p=\infty $ 
at $r >R$) or by a parabolic potential $V_p=\alpha r^2$. The
particles interact through a dipole potential 
$V ({\vec {r}}_i,{\vec {r}}_j)=
q^2/\mid {\vec {r}}_i-{\vec {r}}_j\mid^3$,
where $q^2=\mu _0M^2/4\pi$, or through a screened Coulomb potential 
$V({\vec {r}}_i, {\vec {r}}_j) = (q^2 /
\mid {\vec {r}}_i-{\vec {r}}_j\mid)
\exp (-\kappa {\mid {\vec {r}}_i -{\vec {r}}_j\mid})$,
with $q$ the `particle charge', ${\vec {r}}_i $ is the coordinate
of the $i^{th}$ particle,
and $1/\kappa $ is the screening length where $\kappa =0$ for a Coulomb
cluster and we took $\kappa =2/a_0$ for the screened Coulomb cluster, where 
$a_0$ is the mean inter-particle distance. For given type of inter-particle
interaction and external confinement, only  
two parameters characterize the order of the system: the number of 
particles $N$ and the 
coupling parameter $\Gamma $. We define the characteristic energy of 
inter--particle interaction for dipole clusters 
as $E_0=q^2/a_0^3$ and $E_0=q^2/a_0$ for screened 
Coulomb clusters, where $a_0=2R/N^{1/2}$  for the hard wall 
and $a_0=q^{2/5}\alpha ^{-1/5}$ for parabolic confinement.
In the present calculation we define the coupling parameter 
as $\Gamma =q^2/a_0^3k_BT$ for dipole clusters 
and $\Gamma = (q^2/a_0k_BT)\exp (-\kappa a_0)$ for screened Coulomb clusters.
In \cite {Bubeck} a different dimensionless parameter $\Gamma $ was introduced, where 
$V$ was taken to be the sum over {\it all} pairs of particles.
Our coupling parameter $\Gamma $ is a factor $2.2447$ 
smaller than the one of \cite {Bubeck} for $N=29$. 

The ratio of the particle velocity relaxation time versus the particle
position relaxation time is very small due to the viscosity of 
water and therefore the motion of the particles  
is diffusive. In our simulations we 
will neglect hydrodynamic interactions. 
Following \cite {Ermak} we 
rewrite the stochastic Langevin equations of motion for the position of the particles 
as the equations of motion for Brownian particles 
\begin {equation}
\label{eq1}
\frac {d{\vec {r}}_i }{dt}=
\frac {k_BT}{D_0m}({\sum\limits_{j=1}^{N}}
\frac {dV({\vec {r}}_i,{\vec {r}}_j )}
{d{\vec {r}}}+\frac{dV_p(\vec{r}_i)}{d\vec{r}})+\frac {{\vec F}_L}{m},
\end {equation}
where  $D_0$ is the self-diffusion coefficient, $m$ the particle mass, and
${\vec F}_L$ the randomly fluctuating force acting on the particles due to the 
surrounding media. In the numerical solution of Eq. (1)
we took a time step   
$\Delta t\leq 10^{-4}/(nD_0)$, where $n = N/(\pi R^2)$ is the particle density
and $\Delta t$ was varied within a range  
$(0.02\div 0.05)sec$. The radius of the circular vessel $R=36 \mu m$
and the self-diffusion 
coefficient $D_0=0.35\mu m^2/s$ are taken from the experiment \cite {Bubeck}.  
Following Ref. \cite{Bubeck} we consider dipole clusters consisting of $N=29, 30, 34$ 
particles which have 
different types of packing. In the ordered state the systems of $N=29,30$ 
particles are arranged in a triangular `closed packed' structure 
having, respectively, the shell structure (3:9:17) and (3:9:18) and 
ground-state energy $E=2.2447E_0$ and $E=2.2798E_0$.    
The cluster with $N=34$ particles (4:11:19) $E=2.4198E_0$ has a non-close
packed structure. 
 
We find first the ground-state configuration 
using the Monte-Carlo (MC) technique. Our results for the minimal energy 
configuration coincides exactly with the energy found in \cite {Beda94} for 
Coulomb clusters and in \cite {Belousov98} for 
dipole clusters with parabolic confinement.  
In the experiment \cite {Bubeck} the system was first equilibrated and after 
that the particle trajectories were recorded during $30$ minutes.  
In the simulation we equilibrated the system for about  
$(5\cdot 10^5\div 10^6)$ MC steps after which we started with the 
statistical averaging over time of the different observables.
To obtain reliable results with small statistical error we follow the 
particle trajectories typically during $10^7$ time steps. 

We calculated the time dependence of the mean angular displacement of the 
particles in a specific shell 
$\theta (t)  =  
            {\sum\limits_{i=1}^{N}}(\theta _i(t) - \theta _i(0))/N$,
where $\theta _i(t)$  
is the angular position of the $i^{th}$ particle and $N$ is the number of particles
in the shell.  
In Fig.~1 the angular displacements of the particles in the  first shell (the 
most inner) and the second 
shell are given as function of time for the  cluster of $N=29$ particles.
The angular motion in the third shell is very small because its motion is
hindered by the hard wall.
Notice that 
both the first (thick curve) and the second (thin curve) shells take part in the 
angular motion, but the former rotation is more 
prominent. The inter-shell motion has no preferential direction and with 
time it can be either a clock wise or a counter clock wise rotation. 
With decreasing coupling parameter $\Gamma $ inter-shell rotation becomes 
more pronounced. 

In order to characterize the order of the system we calculate the angular 
diffusion of the particles over a $30min\times 1000$ time interval. The angular 
diffusion coefficient  can be written as
\begin {equation}
\label{eq3}     
D_\theta  =  (<\Delta \theta (t)^2> - <\Delta \theta (t)>^2)/t,
\end {equation}
where $<>$ refers to a time averaging, and the mean relative angular 
displacement rotation of the first shell ($\theta^1(t)$) relative to the
second ($\theta^2(t)$) one is defined
as $\Delta \theta (t) =\theta ^1(t) -\theta ^2(t).$
The radial diffusion coefficient is 
\begin {equation}
\label{eq4}    
\Delta R^2= \frac {1}{N}
\sum\limits_{i=1}^{N}(<r_i(t)^2>-<r_i(t)>^2)/a_0^2,  
\end {equation}
which is a measure of the radial order in the system. 

In Fig.~2 the angular and radial diffusion coefficients are shown for three
different dipole clusters with hard wall 
confinement subjected to the same conditions as in the experiment \cite {Bubeck}.
For the cluster with $N=29$ particles the angular diffusion (solid dots in
Fig.~2(a)) monotonically 
increases with decreasing coupling parameter up to $\Gamma 
\sim 30$ which is a manifestation of angular melting. 
In the interval $\Gamma =10\div 30$ the
inter-shell diffusion remains practically constant, and with further decreasing  
$\Gamma $ it is reduced to about a $20\%$ smaller value.
In the latter region the radial diffusion coefficient starts to rise 
(open dots in Fig.~2(a)), but the cluster retains its 
shell structure. In the range $3<\Gamma <8$ the cluster oscillates between 
the ground state (3:9:17) and 
the metastable state (4:8:17) which leads to a reduction of the angular fluctuations.
Further decreasing the coupling to
$\Gamma \approx 5$ both $D_\theta $ and $\Delta R^2$ rises quickly, indicating
the onset of melting.
A similar qualitative behavior was observed for the dipole cluster with $N=30$ 
particles (see Fig.~2(b)). In the non-close-packed cluster with 
$N=34$ intershell 
rotation occurs over all  $\Gamma $'s considered in the experiment
(see Fig.~2(c)) and no clear regaining of angular order is found in the region
$3<\Gamma <8$. 

In order to obtain further insight into this re-entrant behavior we investigated
the conditions under which this novel effect can be observed.  
Therefore, we varied the following parameters:
1) the viscosity of the medium the particles are moving in;
2) the way of decreasing the coupling of the system (i.e. temperature change 
versus inter-particle interaction strength change);
3) the range of the inter-particle interaction; and  
4) the form of the confinement potential.

To study the melting behavior of the cluster under condition of low viscosity with 
hard wall confinement we performed Langevin molecular dynamics (MD) simulations. 
The Langevin equations for the system of $N$ interacting particles are 
\begin {equation}
\label{eq1}
\frac {d^2{\vec {r}}_i }{dt^2}=
-\nu \frac {d{\vec {r}}_i }{dt}
+\frac {{\vec F}_i}{m}+\frac {{\vec F}_L}{m},
\end {equation}
where  $\nu =D_0m/k_BT$ is the friction and 
${\vec F}_i $ the force from the inter-particle interaction.
As an example, we consider $N=30$ dipole particles moving 
in a medium with a viscosity which is $10^4$ times smaller than the one of water.
Such a low viscosity corresponds to the situation
of colloidal particles moving in a gas with 
pressure 1Pa.
In Fig.~3 the angular 
diffusion coefficient (solid circles) is plotted as function of $\Gamma $. 
Note that now $D_{\theta}$ is about a factor $10^4$ larger as compared to previous
case (see Fig.~2(b)). It is clear that 
changing viscosity does not destroy the re-entrant like behavior.
In fact changing viscosity will not alter the statistical properties
of the system but will only change the time scale for relaxation to equilibrium.

The coupling ($\Gamma$) of the system can be varied in two different ways.
One can decrease 
the inter-particle interaction as was done in the experiment \cite {Bubeck} or  
increase the  temperature of the system in order to induce melting.
The latter approach was used for the $N=30$ cluster and the 
angular diffusion constant is shown in Fig.~3 (open circles).
The inter-particle interaction was chosen equal to the previous one at $\Gamma
= 300$ for $T=300K$. By heating the system no re-entrant melting behavior is
observed and $D_{\theta}$ increases monotonically with increasing temperature
$T \sim 1/\Gamma$. 
Thus re-entrant behavior is only found when the coupling of the 
system is decreased by changing the inter--particle interaction
strength at fixed T.
The reason for this unexpected behavior is that by increasing temperature the
self-diffusion constant, $D_0= k_B T/\nu m$, increases which, because
$D_{\theta} \sim D_0$, implies that the solid dot results in Fig.~3 have to be
multiplied with $T \sim 1/\Gamma$, and the resulting curve (open dots) does not
exhibit a local minimum. The numerical results for the dimensionless variables
(e.g. $D_{\theta}/D_0$) depend only on $\Gamma$ (and N, confinement and
inter-particle interaction) but when converting them into physical variables
$T, \nu, m$ enters. 

Next we investigated whether the type of inter-particle interaction influences  
the occurrence of re-entrant behavior. We consider a cluster with long range
Coulomb interaction ($N=37$ (3:9:25) having the same internal structure as the
$N=30$ dipole cluster) with hard wall confinement using Brownian dynamics (1).
In Fig.~4(a) the angular and radial diffusion 
coefficients are shown as function of $\Gamma $.
Notice that the Coulomb cluster shows  completely different 
melting behavior and $D_\theta $ increases monotonically with decreasing $\Gamma $.
Melting takes place at $\Gamma \sim 40$. 
Next we consider short range inter-particle interaction and as an example we
took the  
screened Coulomb cluster ($\kappa =2/a_0$) with $N=30$ (3:9:18) particles. In Fig.~4(b)
the angular and radial diffusion is shown as function of the coupling.  
$D_\theta $ exhibits a clear re-entrant behavior 
which correlates with an increase of the radial diffusion.
Thus we may conclude that only clusters with short range inter-particle
interaction in a hard wall vessel shows re-entrant behavior.

Last we study the effect of the shape of the confinement potential and consider
the melting behavior of a cluster with short range
inter-particle interaction in a 
parabolic well. 
We choose the $N=25$ dipole cluster (3:9:13).
Fig.~5 shows the angular diffusion and the radius of the 
cluster as function of $\Gamma $. $D_\theta$ clearly does not exhibit a local
minimum, it rises uniformly with decreasing $\Gamma$, and 
melting occurs for $\Gamma \approx 5$, as in the case of hard wall confinement.
The radius of the cluster $R$ and the mean inter-particle distance 
changes proportionally to $\Gamma^{1/2}$.  

In conclusion, we studied the melting transition of 2D clusters with dipole and 
screened Coulomb type of 
interaction confined by a hard wall or a parabolic external potential using 
Brownian dynamics 
simulations. Langevin molecular dynamics simulations were carried out 
in the case of small viscosity. 
We found that only clusters with short range inter-particle interaction 
and confined by a hard wall well exhibit angular freezing before melting,
irrespective of the value of viscosity. 
But it is essential that the coupling of the system is reduced
by decreasing the inter-particle interaction at constant temperature as was done 
in the experiment \cite {Bubeck}. In the other cases, either of Coulomb clusters or 
with parabolic confinement the system shows the usual two step 
melting behavior without any re-entrance.

We showed that re-entrant
behavior is a consequence of the interplay between angular order and radial
oscillations where an increase of the radial fluctuations is able to induce
angular order in clusters with magic number.
With decreasing $\Gamma$, first angular motion sets in, because it
is governed by the lowest energy barriers. Further increasing $\Gamma$ 
leads to an increase of the radial motion/fluctuations which hinder the angular
motion. The later prevents angular motion in case of hard wall confinement.
But for parabolic confinement the average inter-particle distance (see Fig.~5)
decreases 
which results in an equal scaling of the energy barriers for inter-shell and
intra-shell motion. This contrasts with the hard wall confinement case where
the inter-particle distances are
unaltered and the energy barrier for inter-shell jumps decreases leading to an
increase of the radial fluctuations and the inter-shell jumps. 
Thus anharmonic effects are essential for the occurrence of this
re-entrant behavior. Anharmonicity is enhanced in systems with hard wall confinement
and for short-range inter-particle interaction.

One of us (FMP) thanks C. Bechinger for fruitful discussions. This work is supported 
by the Flemish Science Foundation (FWO-Vl), IUAP-IV and the Russian 
Foundation for Fundamental Investigation 
96--023--9134a. One of us (FMP) is a Researcher 
Director with FWO-Vl and I.V.S and V.A.S are supported by a DWTC--fellowship.

\begin {center} { FIGURES } \end {center}

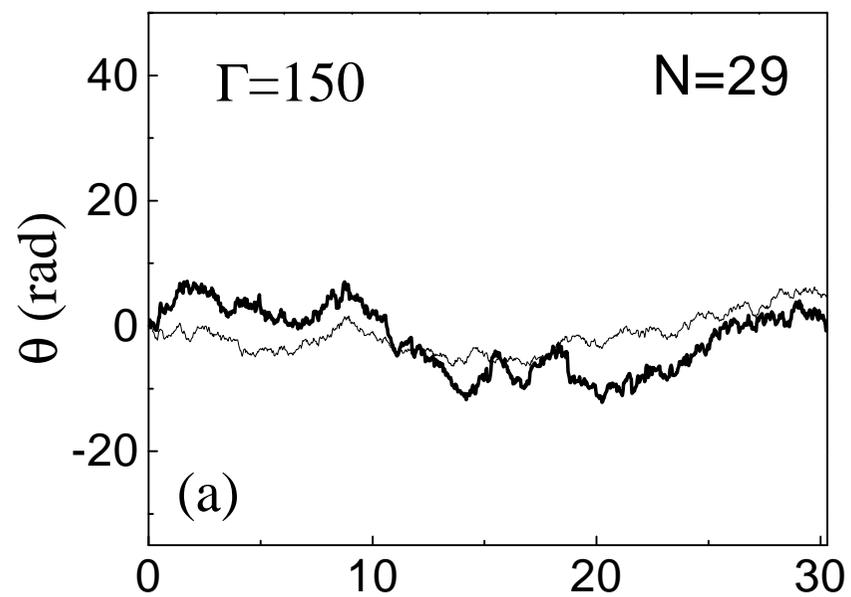
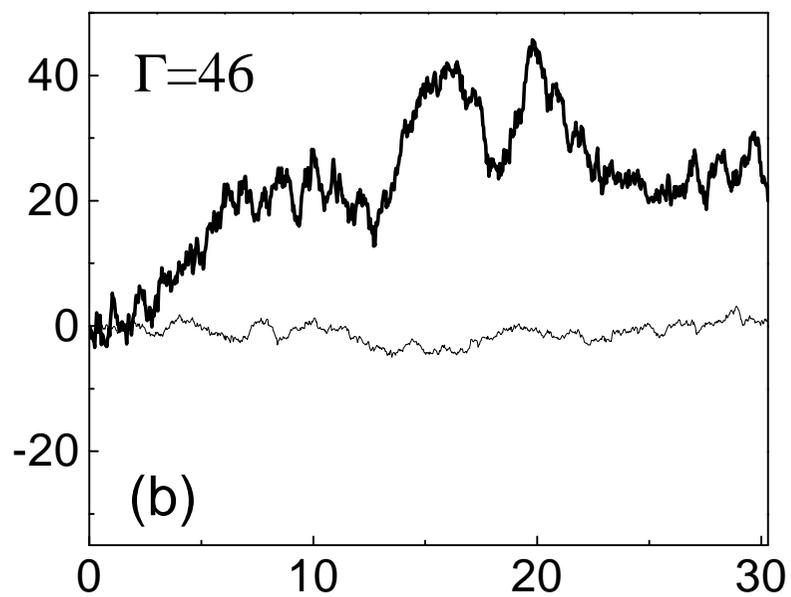
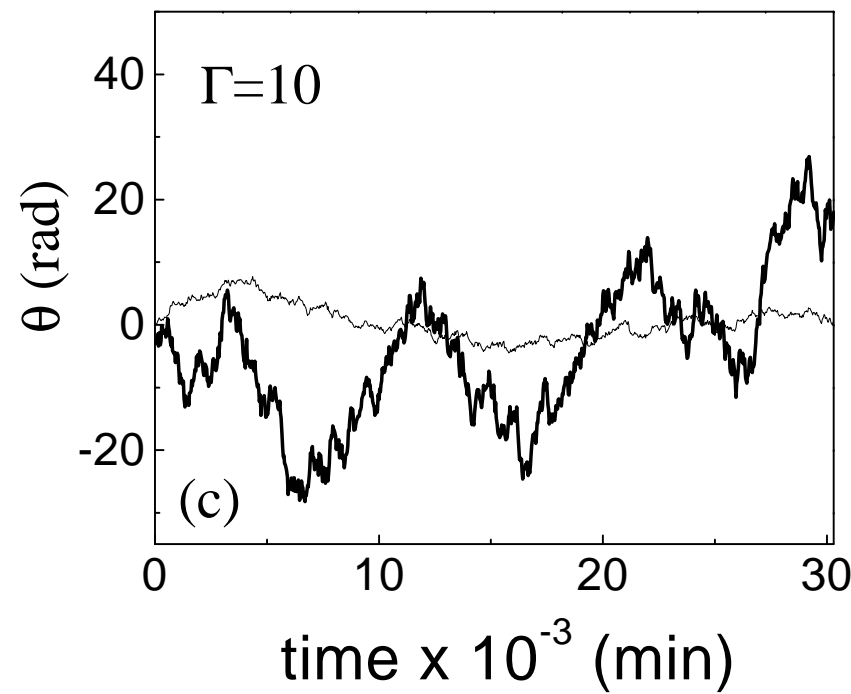
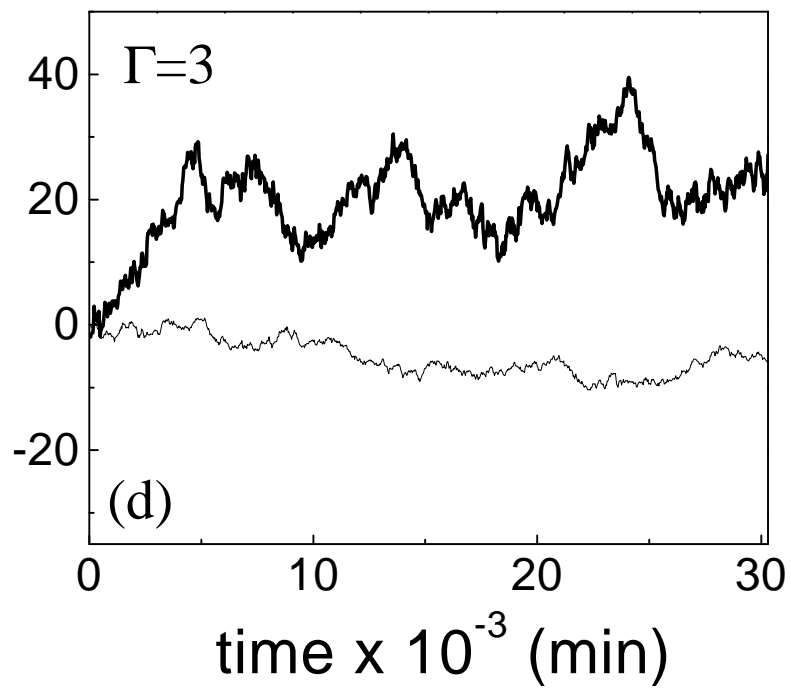
FIG. 1. The angular displacement of the particles on the first (thick curves) 
and second (thin curves) shell for a cluster with 
$N=29$ particles for different values of the coupling parameter: (a) $\Gamma = 150$,
(b) $\Gamma = 46$, (c) $\Gamma =10$, and (d) $\Gamma =3$.

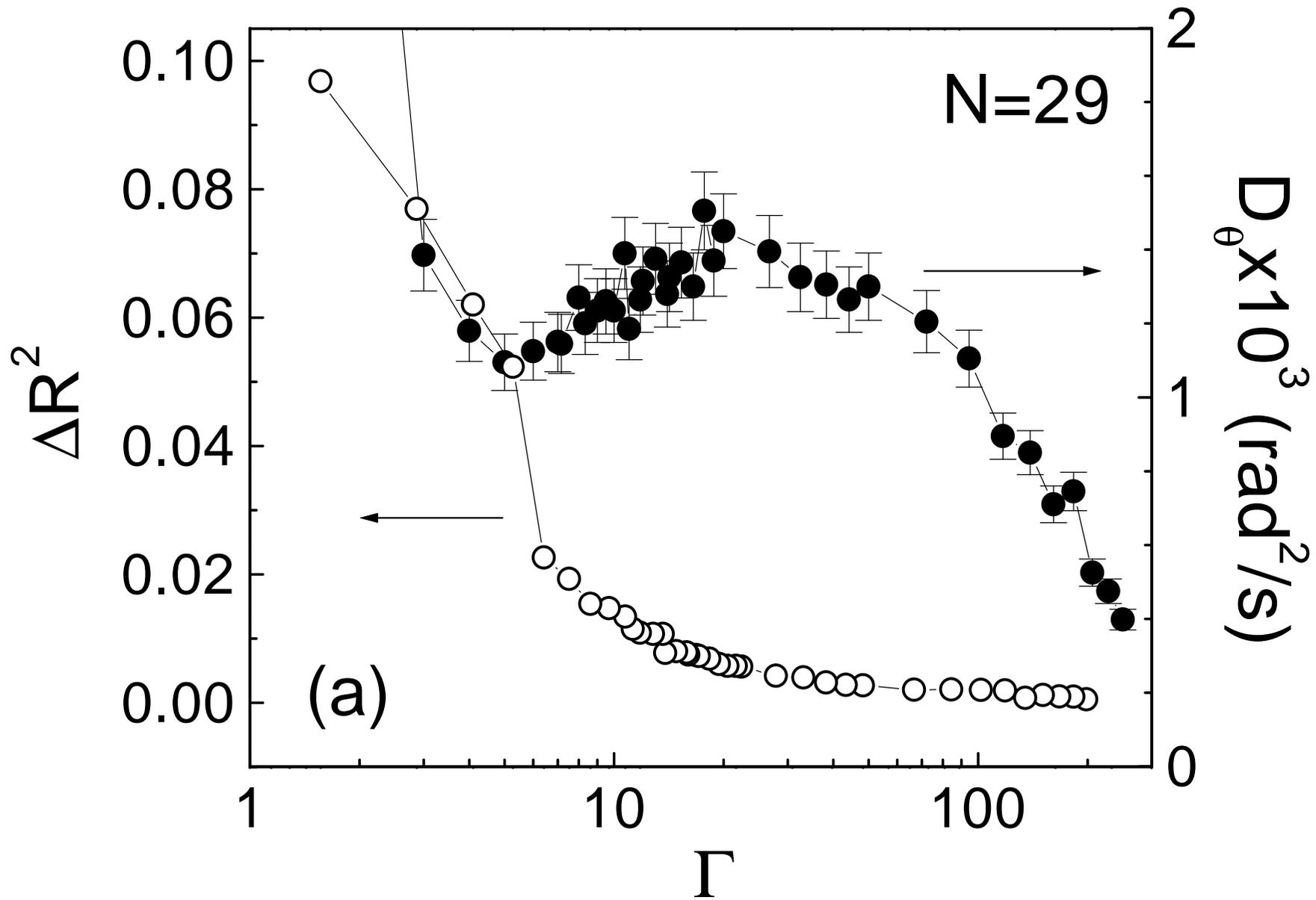
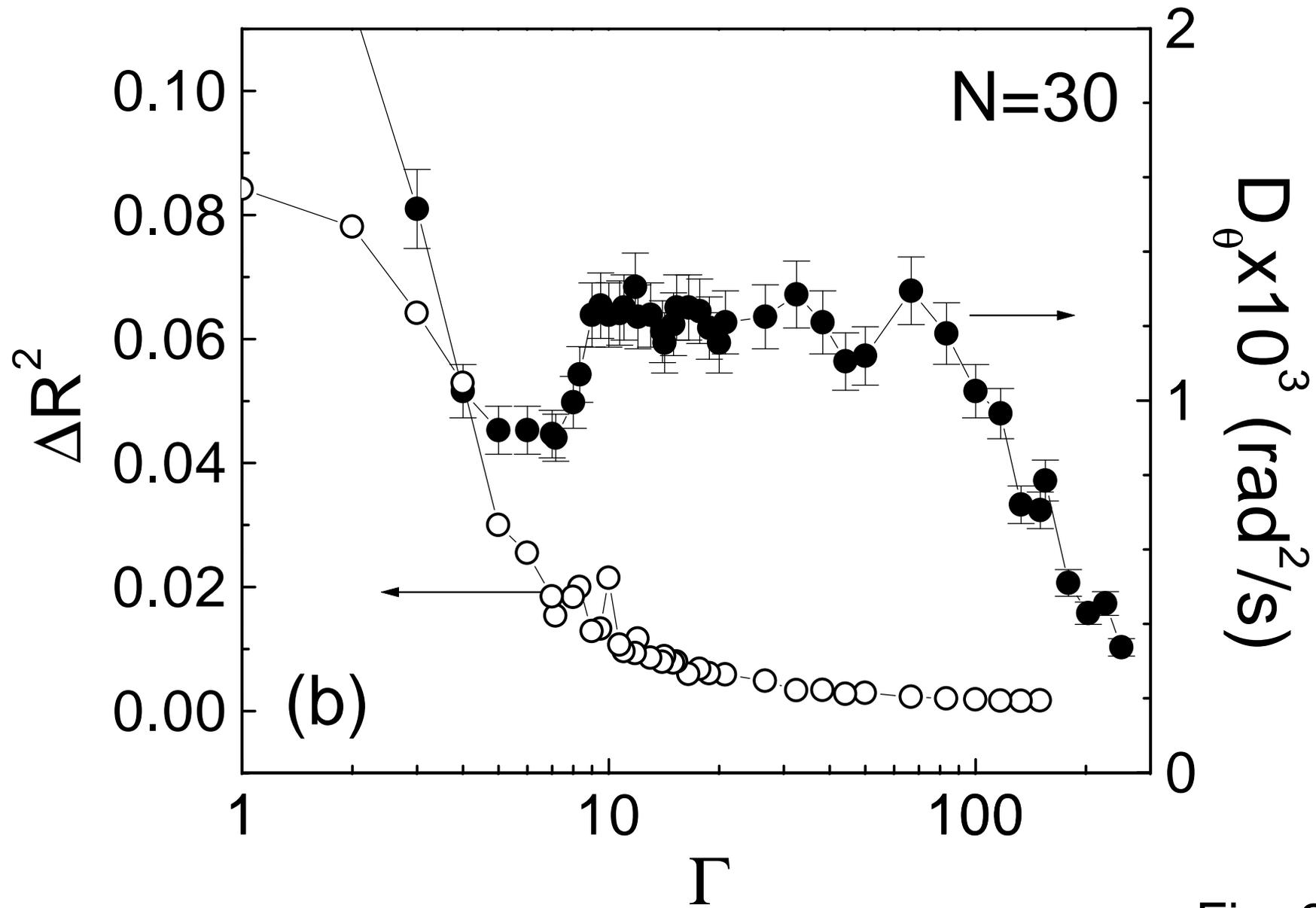
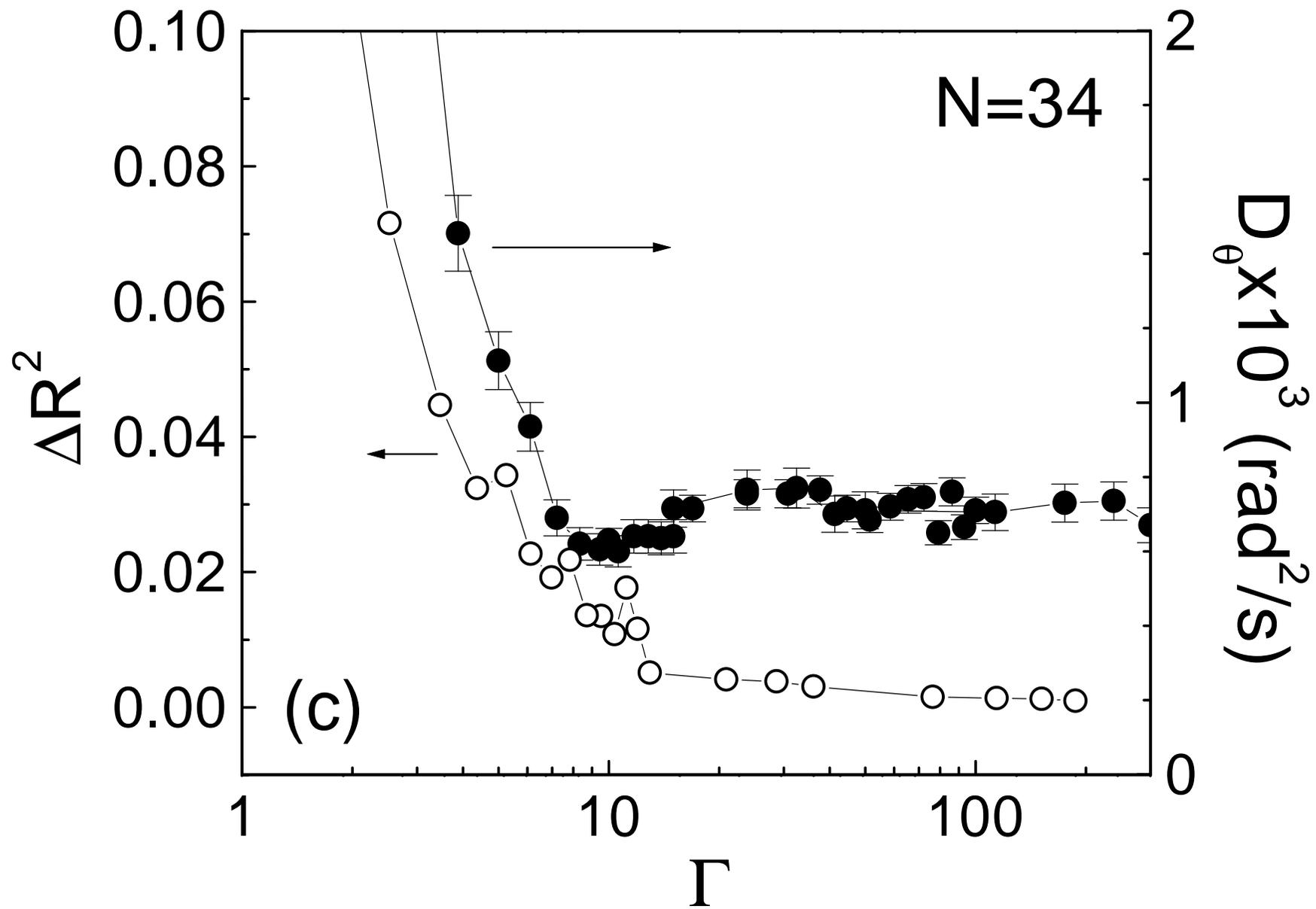
FIG. 2. The angular diffusion  $D_\theta $ (solid circles) and the radial diffusion
($\Delta R^2$)  (open circles) coefficients
as function of $\Gamma $ for clusters with (a) $N=29$, (b) 
$N=30$, and (c) $N=34$ particles. 

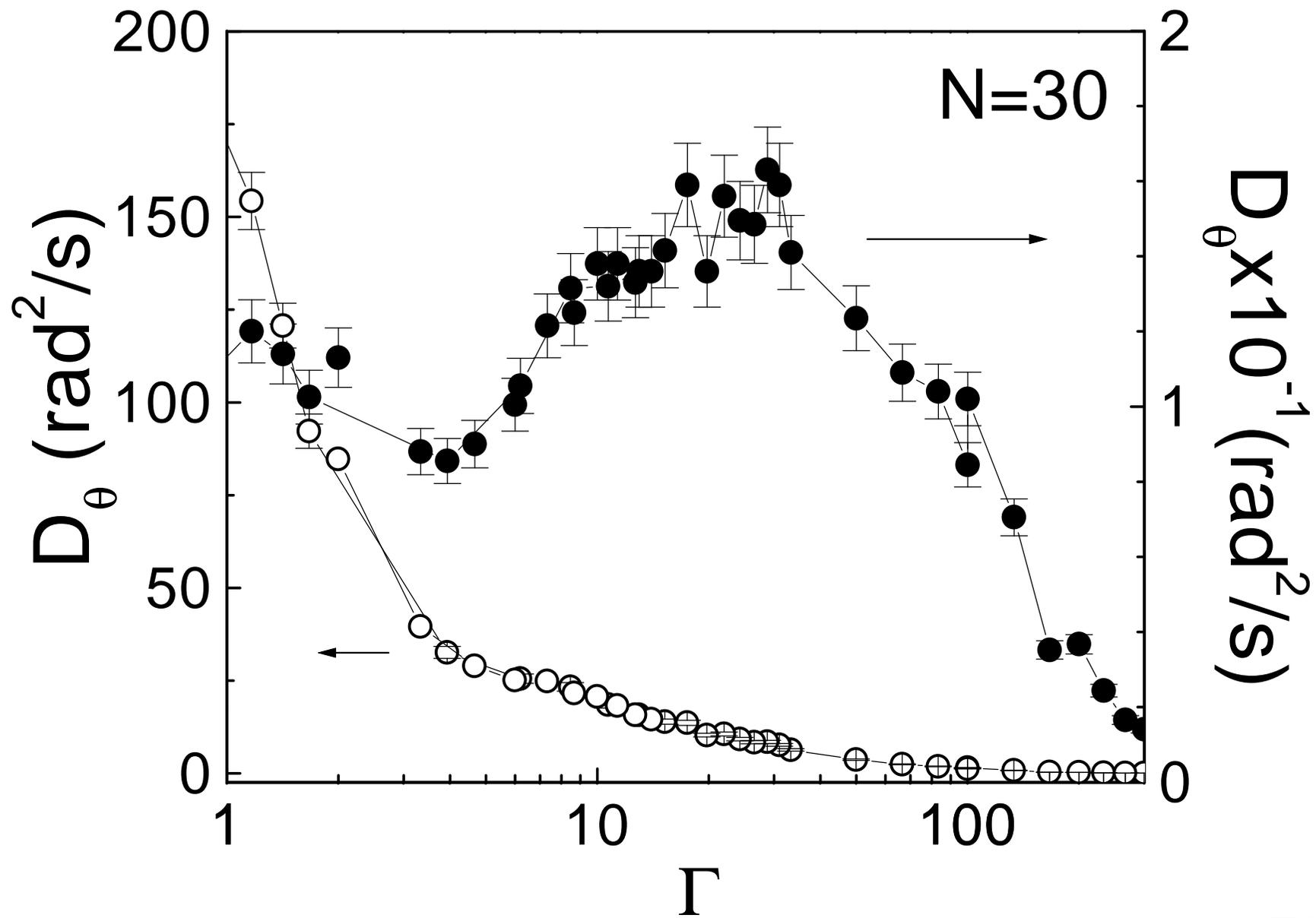
FIG. 3. The angular diffusion coefficients as a function of the strength
of the inter-particle interaction at constant temperature (solid circles) and
as function of temperature ($\Gamma \sim 1/T$) for fixed inter-particle
interaction strength (open 
circles) for a dipole cluster with $N=30$ in the case of small viscosity.
 
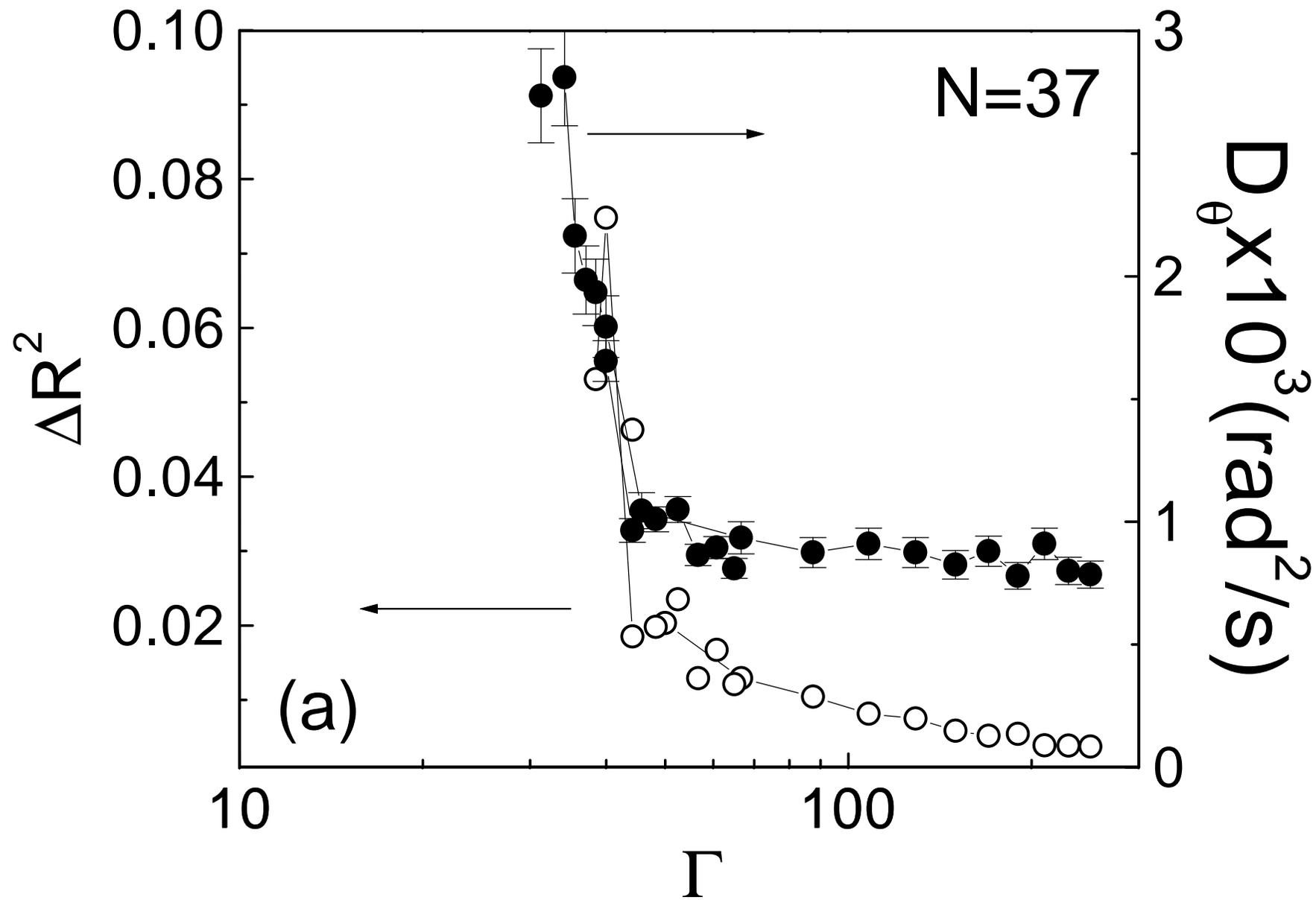
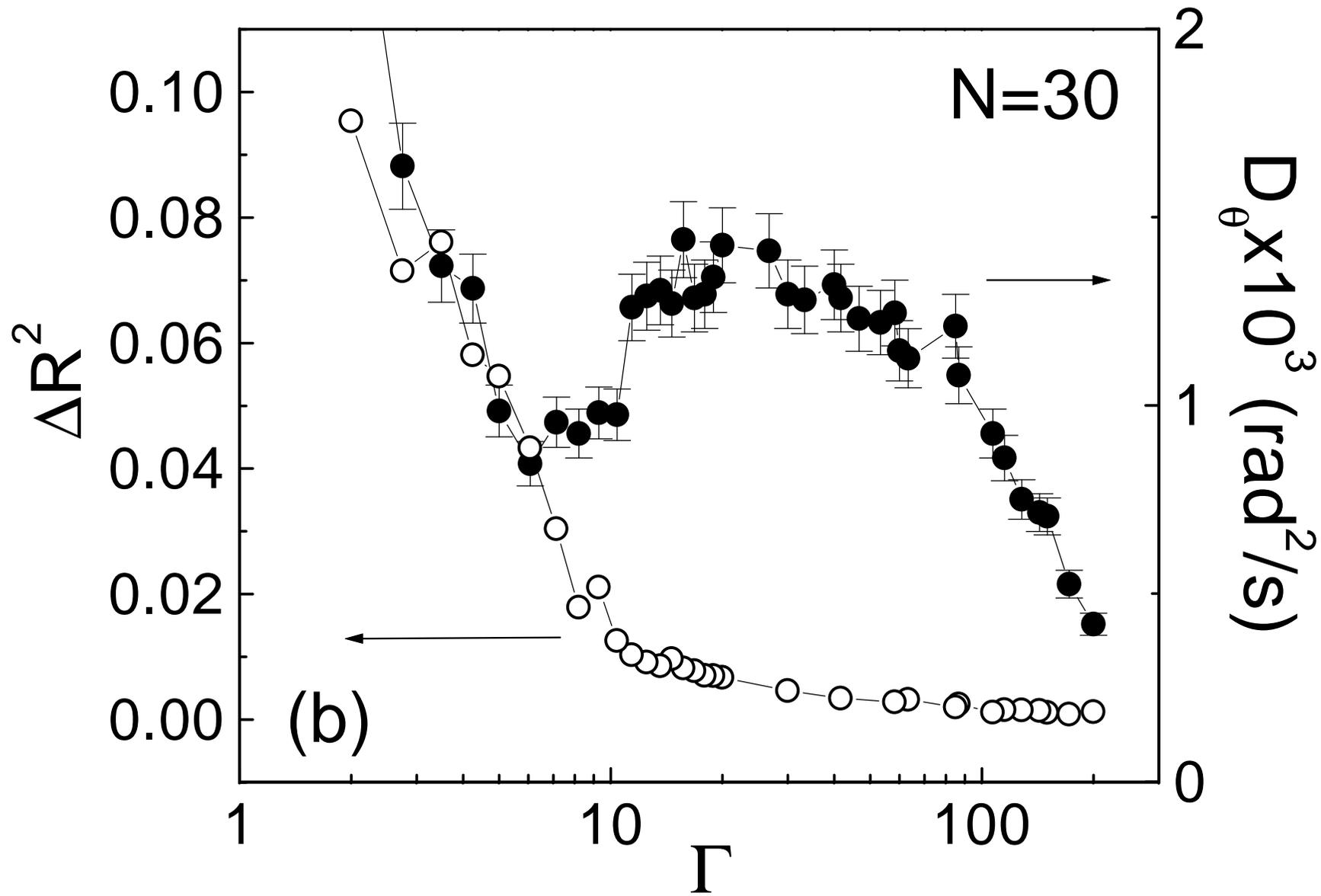
FIG. 4. The angular (solid circles) and radial diffusion (open circles)
coefficients as function of $\Gamma $ for: (a) the Coulomb cluster with $N=37$ 
particles, and (b) the screened Coulomb cluster with $N=30$ particles confined
by a hard wall potential.

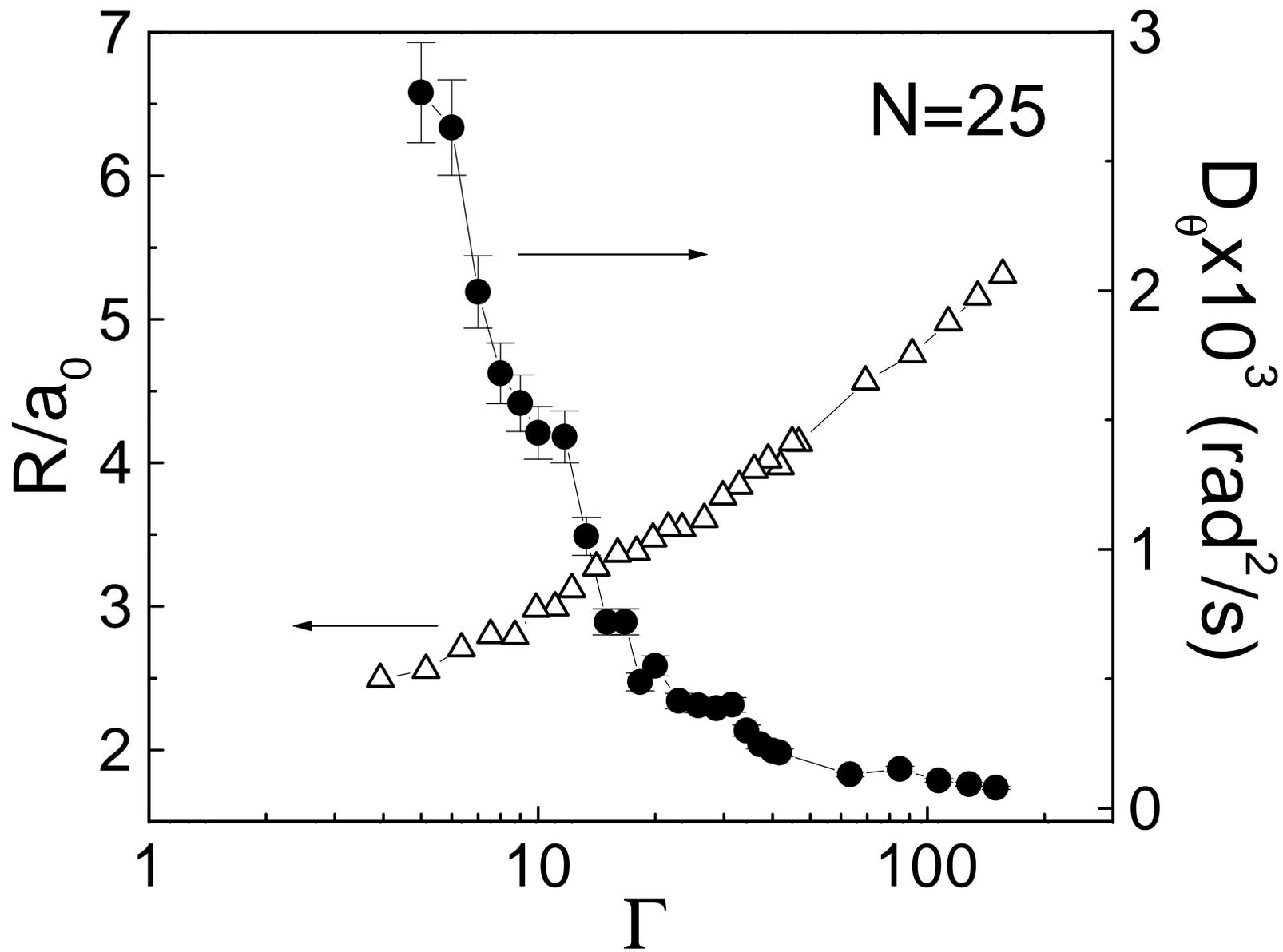
FIG. 5. The angular diffusion coefficient (solid circles) and radius of the cluster  
(open triangles) as function of $\Gamma$ for a $N=25$ dipole cluster
in a parabolic well.

\newpage
\end {document}